\documentclass[aps,prl,preprint,tightenlines,superscriptaddress,showpacs,byrevtex]{revtex4}

\def \ks {K_{\rm S}^0}
\usepackage{graphicx}
\usepackage{color}

\begin{document}


\preprint{\vbox{ \hbox{BELLE-CONF-0507}
                 \hbox{LP2005-142}
                 \hbox{EPS05-479} 
}}

\title{ \quad\\[0.5cm] Search for Lepton Flavor Violating $\tau^-$ Decays
{with a}
$\ks$ Meson}

\affiliation{Aomori University, Aomori}
\affiliation{Budker Institute of Nuclear Physics, Novosibirsk}
\affiliation{Chiba University, Chiba}
\affiliation{Chonnam National University, Kwangju}
\affiliation{University of Cincinnati, Cincinnati, Ohio 45221}
\affiliation{University of Frankfurt, Frankfurt}
\affiliation{Gyeongsang National University, Chinju}
\affiliation{University of Hawaii, Honolulu, Hawaii 96822}
\affiliation{High Energy Accelerator Research Organization (KEK), Tsukuba}
\affiliation{Hiroshima Institute of Technology, Hiroshima}
\affiliation{Institute of High Energy Physics, Chinese Academy of Sciences, Beijing}
\affiliation{Institute of High Energy Physics, Vienna}
\affiliation{Institute for Theoretical and Experimental Physics, Moscow}
\affiliation{J. Stefan Institute, Ljubljana}
\affiliation{Kanagawa University, Yokohama}
\affiliation{Korea University, Seoul}
\affiliation{Kyoto University, Kyoto}
\affiliation{Kyungpook National University, Taegu}
\affiliation{Swiss Federal Institute of Technology of Lausanne, EPFL, Lausanne}
\affiliation{University of Ljubljana, Ljubljana}
\affiliation{University of Maribor, Maribor}
\affiliation{University of Melbourne, Victoria}
\affiliation{Nagoya University, Nagoya}
\affiliation{Nara Women's University, Nara}
\affiliation{National Central University, Chung-li}
\affiliation{National Kaohsiung Normal University, Kaohsiung}
\affiliation{National United University, Miao Li}
\affiliation{Department of Physics, National Taiwan University, Taipei}
\affiliation{H. Niewodniczanski Institute of Nuclear Physics, Krakow}
\affiliation{Nippon Dental University, Niigata}
\affiliation{Niigata University, Niigata}
\affiliation{Nova Gorica Polytechnic, Nova Gorica}
\affiliation{Osaka City University, Osaka}
\affiliation{Osaka University, Osaka}
\affiliation{Panjab University, Chandigarh}
\affiliation{Peking University, Beijing}
\affiliation{Princeton University, Princeton, New Jersey 08544}
\affiliation{RIKEN BNL Research Center, Upton, New York 11973}
\affiliation{Saga University, Saga}
\affiliation{University of Science and Technology of China, Hefei}
\affiliation{Seoul National University, Seoul}
\affiliation{Shinshu University, Nagano}
\affiliation{Sungkyunkwan University, Suwon}
\affiliation{University of Sydney, Sydney NSW}
\affiliation{Tata Institute of Fundamental Research, Bombay}
\affiliation{Toho University, Funabashi}
\affiliation{Tohoku Gakuin University, Tagajo}
\affiliation{Tohoku University, Sendai}
\affiliation{Department of Physics, University of Tokyo, Tokyo}
\affiliation{Tokyo Institute of Technology, Tokyo}
\affiliation{Tokyo Metropolitan University, Tokyo}
\affiliation{Tokyo University of Agriculture and Technology, Tokyo}
\affiliation{Toyama National College of Maritime Technology, Toyama}
\affiliation{University of Tsukuba, Tsukuba}
\affiliation{Utkal University, Bhubaneswer}
\affiliation{Virginia Polytechnic Institute and State University, Blacksburg, Virginia 24061}
\affiliation{Yonsei University, Seoul}
  \author{K.~Abe}\affiliation{High Energy Accelerator Research Organization (KEK), Tsukuba} 
  \author{K.~Abe}\affiliation{Tohoku Gakuin University, Tagajo} 
  \author{I.~Adachi}\affiliation{High Energy Accelerator Research Organization (KEK), Tsukuba} 
  \author{H.~Aihara}\affiliation{Department of Physics, University of Tokyo, Tokyo} 
  \author{K.~Aoki}\affiliation{Nagoya University, Nagoya} 
  \author{K.~Arinstein}\affiliation{Budker Institute of Nuclear Physics, Novosibirsk} 
  \author{Y.~Asano}\affiliation{University of Tsukuba, Tsukuba} 
  \author{T.~Aso}\affiliation{Toyama National College of Maritime Technology, Toyama} 
  \author{V.~Aulchenko}\affiliation{Budker Institute of Nuclear Physics, Novosibirsk} 
  \author{T.~Aushev}\affiliation{Institute for Theoretical and Experimental Physics, Moscow} 
% ...
\collaboration{The Belle Collaboration}

\begin{abstract}
We have searched for 
the $\tau$ lepton flavor violating {decays}
$\tau^-\rightarrow \ell^-\ks$ ($\ell = e \mbox{ or } \mu$),
using a data sample of
281 fb$^{-1}$ collected with
the Belle detector at the KEKB $e^+e^-$ asymmetric-energy collider.
No evidence for a signal {was} found
in either of the decay modes,
{and we} set the following upper limits 
for the branching {fractions:}
${\cal{B}}(\tau^-\rightarrow e^-\ks) < 5.6\times 10^{-8}$
and 
${\cal{B}}(\tau^-\rightarrow \mu^-\ks) < 4.9\times 10^{-8}$ 
at the 90\% confidence level. 
{These results are {improvements}
by factors of 16 and 19, respectively,
compared with previously published limits from CLEO.
}

\end{abstract}


\maketitle

\section{Introduction}

{Lepton flavor violation (LFV) 
{is allowed} in many extensions of the Standard Model (SM),
{such as} Supersymmetry (SUSY) and leptoquark models.}
Lepton flavor violating decays with $\ks$ mesons occur in models
{with either heavy singlet Dirac neutrinos~\cite{cite:amon}, 
$R-$parity violation in SUSY~\cite{cite:rpv}
or
dimension-six effective fermionic operators that induce $\tau-\mu$ mixing~\cite{cite:six_fremionic}.}
Experiments at the  $B$-factories
allow 
searches
for such {decays} with
a very high sensitivity.
The best upper limits {of}
${\cal{B}}(\tau^-\rightarrow e^-\ks) < 9.1\times 10^{-7}$
and 
${\cal{B}}(\tau^-\rightarrow \mu^-\ks) < 9.5\times 10^{-7}$
at the 90\% confidence level
were obtained 
in  the CLEO experiment
using 13.9 fb${}^{-1}$ of data~\cite{cite:cleo}.

In this paper,
we {report on  a} {search} for
{LFV $\tau$ decays}
$\tau^-\rightarrow \ell^-\ks$
($\ell = e \mbox{ or } \mu$)\footnotemark[2]
with a data sample
of 281 fb$^{-1}$ 
collected at the $\Upsilon(4S)$ resonance
and 60 MeV below it
with the Belle detector at the KEKB  $e^+e^-$ 
asymmetric-energy collider~\cite{kekb}. 
\footnotetext[2]{Unless otherwise stated, charge 
conjugate decays are implied throughout
this paper.}

The Belle detector is a large-solid-angle magnetic spectrometer that
consists of a silicon vertex detector (SVD), 
a 50-layer central drift chamber (CDC), 
an array of aerogel threshold \v{C}erenkov counters (ACC), a barrel-like arrangement of 
time-of-flight scintillation counters (TOF), and an electromagnetic calorimeter 
comprised of CsI(Tl) {crystals (ECL), all located} inside
a superconducting solenoid coil
that provides a 1.5~T magnetic field.  
An iron flux-return located outside of the coil is instrumented to detect $K_{\rm{L}}^0$ mesons 
and to identify muons (KLM).  
The detector is described in detail elsewhere~\cite{Belle}.

{Particle identification 
is very important in this measurement.
Particle identification likelihood variables are based on}
the ratio of the energy 
deposited in the ECL to the momentum measured in the SVD and CDC, 
the shower shape in the ECL, 
the particle range in the KLM, 
the hit information from the ACC,
{the measured $dE/dX$} in the CDC 
and {the particle's time-of-flight} from the TOF.
{For lepton identification,
we use  a likelihood ratio based on the}
electron probability ${\cal P}(e)$~\cite{EID} and 
muon probability ${\cal P}({\mu})$~\cite{MUID} 
determined by
the responses of the appropriate subdetectors.

{For Monte Carlo (MC) studies,}
the following programs have been used to
generate background events:
KORALB/TAUOLA~\cite{cite:koralb_tauola} for $\tau^+\tau^-$, 
QQ~\cite{cite:qq} for $B\bar{B}$ and continuum,
BHLUMI~\cite{BHLUMI} for {Bhabha events,}
KKMC~\cite{KKMC} for $e^+e^-\rightarrow\mu^+\mu^-$ and
AAFH~\cite{AAFH} for two-photon processes.
Signal MC is generated by KORALB/TAUOLA.
{Signal $\tau$ decays are  two-body 
and assumed}
to  have a uniform angular distribution
{in the $\tau$} lepton's rest frame. 
The Belle detector response is simulated by a GEANT 3~\cite{cite:geant3} 
based program.
{All kinematic variables are
calculated in the laboratory frame
unless stated otherwise.
In particular,
variables
calculated in the $e^+e^-$ center-of-mass (CM) frame 
are indicated by the superscript ``CM''.}


%
%




\section{Data Analysis}

We search for $\tau^+\tau^-$ events
in which one $\tau$ (signal side) decays
into $\ell\ks$ where {$\ell$ is $e$ or $\mu$} 
and $\ks$ decays into $\pi^+\pi^-$,
{while} 
the other $\tau$ (tag side) decays into one charged particle 
{of opposite sign to the lepton}
with any number of
additional photons and neutrinos. 
Thus, the experimental signature is:
\begin{center}
$\left\{
\tau^- \rightarrow \ell^-(=e^-\mbox{ or }\mu^-) + \ks(\rightarrow\pi^+\pi^-)
\right\} 
~+
~ \left\{ \tau^+ \rightarrow ({\rm a~track})^+ + (n^{\rm TAG}_{\gamma} \ge 0)
 + X(\rm{missing}) 
\right\}$.
\end{center}
{All charged tracks} and photons 
are required to be reconstructed within the fiducial volume 
defined by $-0.866 < \cos\theta < 0.956$,
where $\theta$ is the polar angle with
respect to the direction opposite to the $e^+$ beam.
Charged tracks should have
momentum transverse to the $e^+$ beam
$p_t > 0.1$ GeV/$c$ 
and 
photons should have energies
$E_{\gamma} > 0.1$ GeV.

{Signal events are isolated with following selections.}
We first demand that the four tracks have zero net charge.  
The
magnitude of the thrust~\cite{thrust} is required to be larger than 0.9 to suppress
the $q\bar{q}$ continuum background.  The event {must} have a 3-1
prong configuration relative to the plane perpendicular to the thrust
axis.  
The $\ks$ is 
reconstructed 
from two 
{oppositely-charged pions}
that
have an invariant mass within  {0.482 GeV/$c^2 < M_{\pi^+\pi^-} < 0.514$ {GeV/$c^2$}.}
The $\pi^+\pi^-$ vertex is required to
be displaced from the interaction point (IP)
in the direction of the pion pair momentum~\cite{cite:ks}.
{In order to avoid fake $\ks$ candidates 
from  {photon} conversions
(i.e. $\gamma \rightarrow e^+e^-$),}
the invariant mass reconstructed 
by assigning the electron mass to the tracks,
is required to be greater than 0.2 GeV/$c^2$.
{We apply the lepton identification requirements to each track 
except for the two tracks that are part of the $\ks$ candidate
on the signal side.}
{The electron and muon {identification} criteria are}
${\cal P}(e) > 0.9$ with $p > 0.3$ GeV/$c$
and 
${\cal P}(\mu) > 0.9$ with $p > 0.6$ GeV/$c$,
respectively.
After the event selection described above, 
{most of the remaining background} comes from
$\tau^+\tau^-$ and continuum  events 
{including a single $\ks$ meson.}

\begin{figure}[h]
\begin{center}
 \resizebox{0.7\textwidth}{0.6\textwidth}{\includegraphics
 {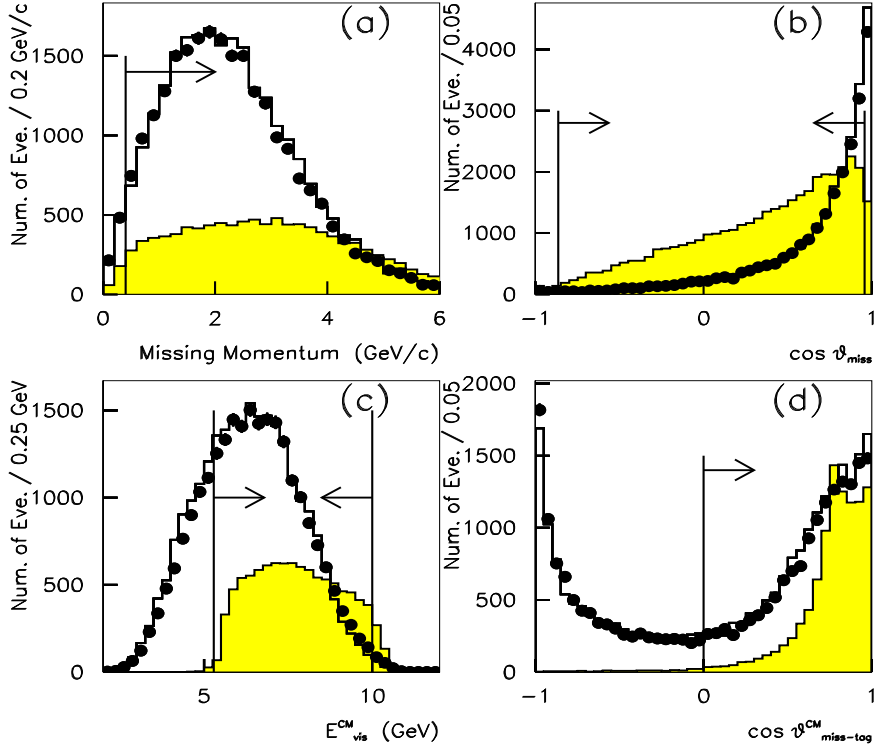}}
 \caption{ 
 Kinematical distributions used in the event selection
 after $\ks$ mass and muon {identification requirements:}
 (a) the momentum of the missing particle;
 (b) the polar angle of the missing particle; 
 (c) the total visible energy in the CM frame;
 (d) the opening angle between the missing particle and
 tag-side track in the CM frame.
 The signal MC distributions are indicated by the filled histograms, 
 {the total background} including 
 $\tau^+\tau^-$ and $q\bar{q}$ {is shown by} the open histogram, 
 and closed circles are data.
 While the signal MC ($\tau^-\rightarrow\mu^-\ks$) 
 distribution is normalized arbitrarily, 
 {the data and background MC} are normalized to the same luminosity.
 {Selected regions are indicated  by arrows from the marked cut {boundaries.}} 
}
\label{fig:cut}
\end{center}
\end{figure}

{To ensure that the missing particles are neutrinos rather
than photons or charged particles that are outside the detector
acceptance, we impose additional requirements on the missing
momentum vector, $\vec{p}_{\rm miss}$, 
calculated by subtracting the
vector sum of the momenta}
of
all tracks and photons 
from the sum of the $e^+$ and $e^-$ beam momenta.
{We require that the magnitude of $\vec{p}_{\rm miss}$ 
be greater than
0.4 GeV/c and that it point into the fiducial volume of the
detector:}
$-0.866 < \cos\theta_{\rm{miss}} < 0.956$,
as shown for {the} $\tau^-\rightarrow\mu^-\ks$ mode 
in Fig.~\ref{fig:cut} (a) and (b).
The total visible energy in the CM frame  $E^{\mbox{\rm{\tiny{CM}}}}_{\rm{vis}}$,
is defined as the sum of the energies
{of the $\ks$ candidate,
the lepton,
the tag-side track (with pion mass assumed) 
and all photon candidates.
We require it to satisfy}
$5.29$ GeV $< E^{\mbox{\rm{\tiny{CM}}}}_{\rm{vis}} < 10.0$ GeV 
(see Fig.~\ref{fig:cut} (c)).
{Since neutrinos are emitted only on the tag-side,
the direction of 
{$\vec{p}_{\rm miss}$
should lie within the tag-side of the event.}}
The cosine of the
opening angle between 
{$\vec{p}_{\rm miss}$}
and
{the} tag-side track 
in the CM system,
$\cos \theta^{\mbox{\rm \tiny CM}}_{\rm tag-miss}$
{is therefore required to be greater than 0}
(see Fig.~\ref{fig:cut} (d)).
To suppress the continuum background,
the following requirements 
on
the number of the photon candidates on the signal and tag side
{are imposed:}
$n_{\gamma}^{\rm{SIG}}\leq 1$ and $n_{\gamma}^{\rm{TAG}}\leq 2$,
respectively.

%
%
%

\begin{figure}[t]
\begin{center}
 \resizebox{.7\textwidth}{!}{\includegraphics
 {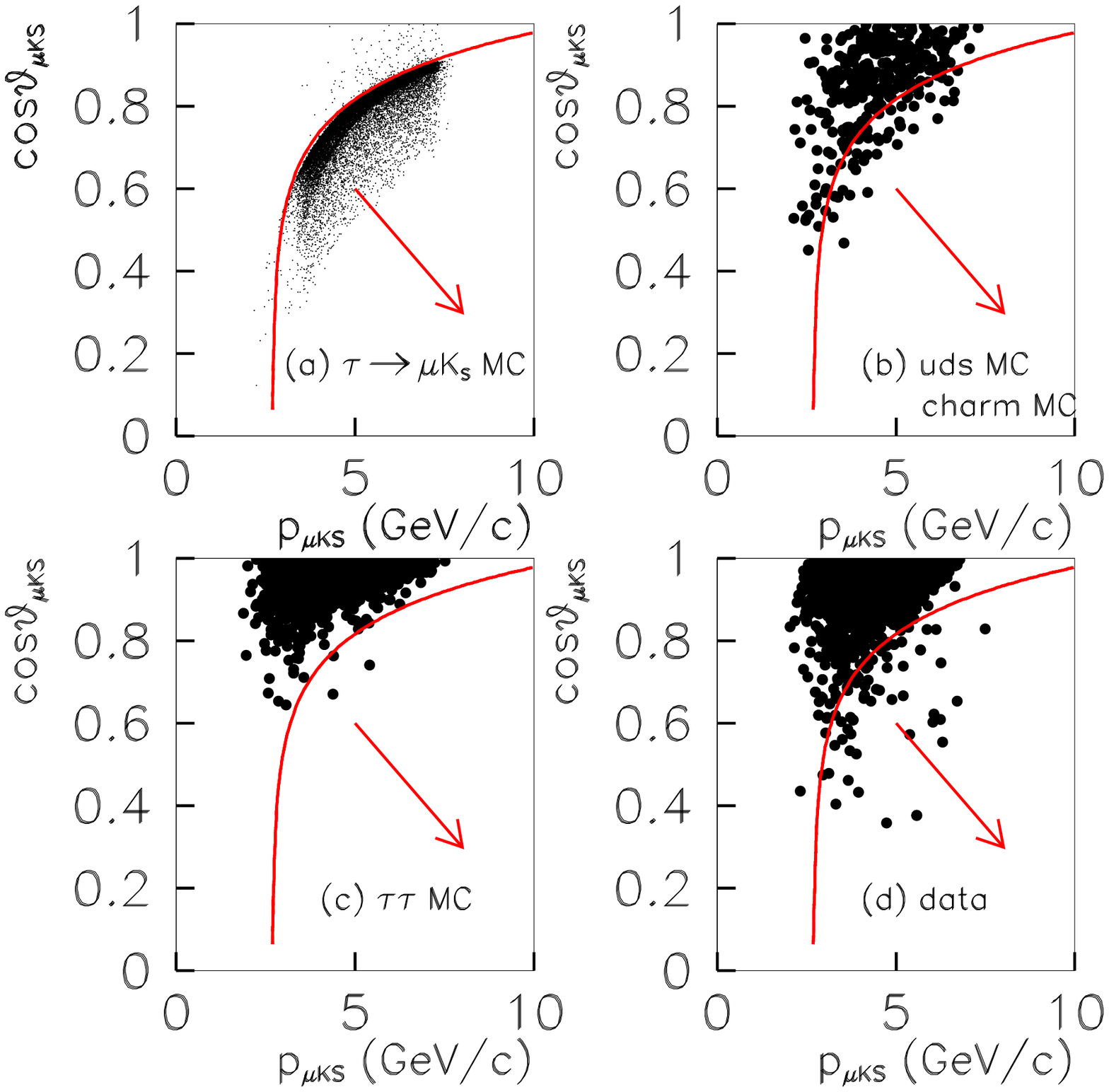}}
 \caption{
Scatter-plots 
 of (a) signal MC ($\tau^-\rightarrow\mu^-\ks$),  (b) continuum MC,  
(c) generic $\tau^+\tau^-$ MC events 
 and (d) data on {the}
$p_{\ell K_{\rm S}}$ vs $\cos \theta_{\ell K_{\rm S}}$ plane.
 Selected regions are indicated by curves with arrows.
 \label{fig:pmiss_vs_mmiss2}
 }
 \end{center}
\end{figure}

Finally, 
the correlation between {the}
reconstructed momentum of {the} $\ell\ks$ system,
$p_{\ell K_{\rm S}}$,
and
the cosine of the opening angle
{between the lepton and $\ks$,}
$\cos \theta_{\ell K_{\rm S}}$
{is employed to exclude 
background from generic $\tau^+\tau^-$
and continuum:}
$\cos \theta_{\ell K_{\rm S}} < 0.14\times\log(p_{\ell K_{\rm S}}-2.7)+0.7$,
where $p_{\ell K_{\rm S}}$ is in GeV/$c$
(see Fig. \ref{fig:pmiss_vs_mmiss2}). 
While this condition retains
99\% of the signal events,
99\% of the generic $\tau^+\tau^-$ and 84\% of {the} 
$uds$ continuum background
are removed.
The signal detection efficiencies 
for {the} {$\tau^-\rightarrow e^-\ks$} and
{$\tau^-\rightarrow\mu^-\ks$} modes are
15.0\% and 16.2\%
after all criteria applied, respectively.

\section{Results}

Signal candidates are examined in the two-dimensional 
space of the $\ell^-\ks$ invariant 
mass, $M_{\rm {inv}}$, and the difference of their energy from the 
beam energy in the CM system, $\Delta E$.
A signal event should have $M_{\rm {inv}}$
close to the $\tau$-lepton mass
and 
$\Delta E$ close to 0.
{For both modes,
the $M_{\rm {inv}}$ and $\Delta E$  resolutions} are parameterized 
from the MC distributions around the peak  
{{with  bifurcated Gaussian shapes}
to account for initial state radiation.
These have widths 
$\sigma^{\rm{high/ low}}_{M_{\rm{inv}}} = 6.2/ 7.4$ MeV/$c$$^2$ and 
$\sigma^{\rm{high/ low}}_{\Delta E} = 20/ 26$ MeV,
for the $\tau^-\rightarrow e^-\ks$ mode
{and,}
$\sigma^{\rm{high/ low}}_{M_{\rm{inv}}} = 6.1/ 5.9$ MeV/$c$$^2$ and 
$\sigma^{\rm{high/ low}}_{\Delta E} = 19/ 23$ MeV,
respectively for the $\tau^-\rightarrow \mu^-\ks$ mode,}
where the ``high/low'' superscript indicates the higher/lower side 
of the peak.

\begin{figure}[t]
\begin{center}
 \resizebox{0.45\textwidth}{0.45\textwidth}{\includegraphics
 {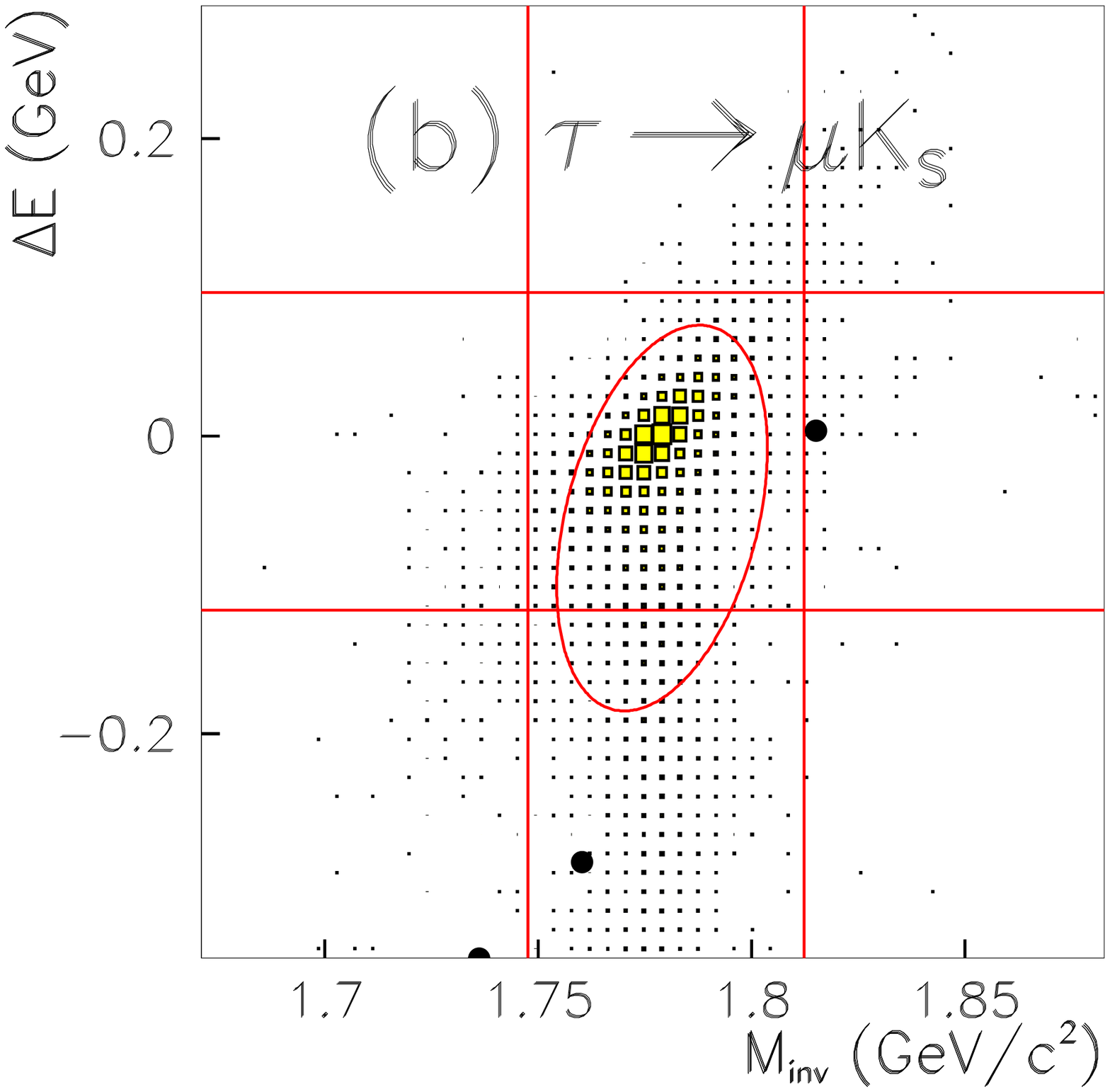}}
 \resizebox{0.45\textwidth}{0.45\textwidth}{\includegraphics
 {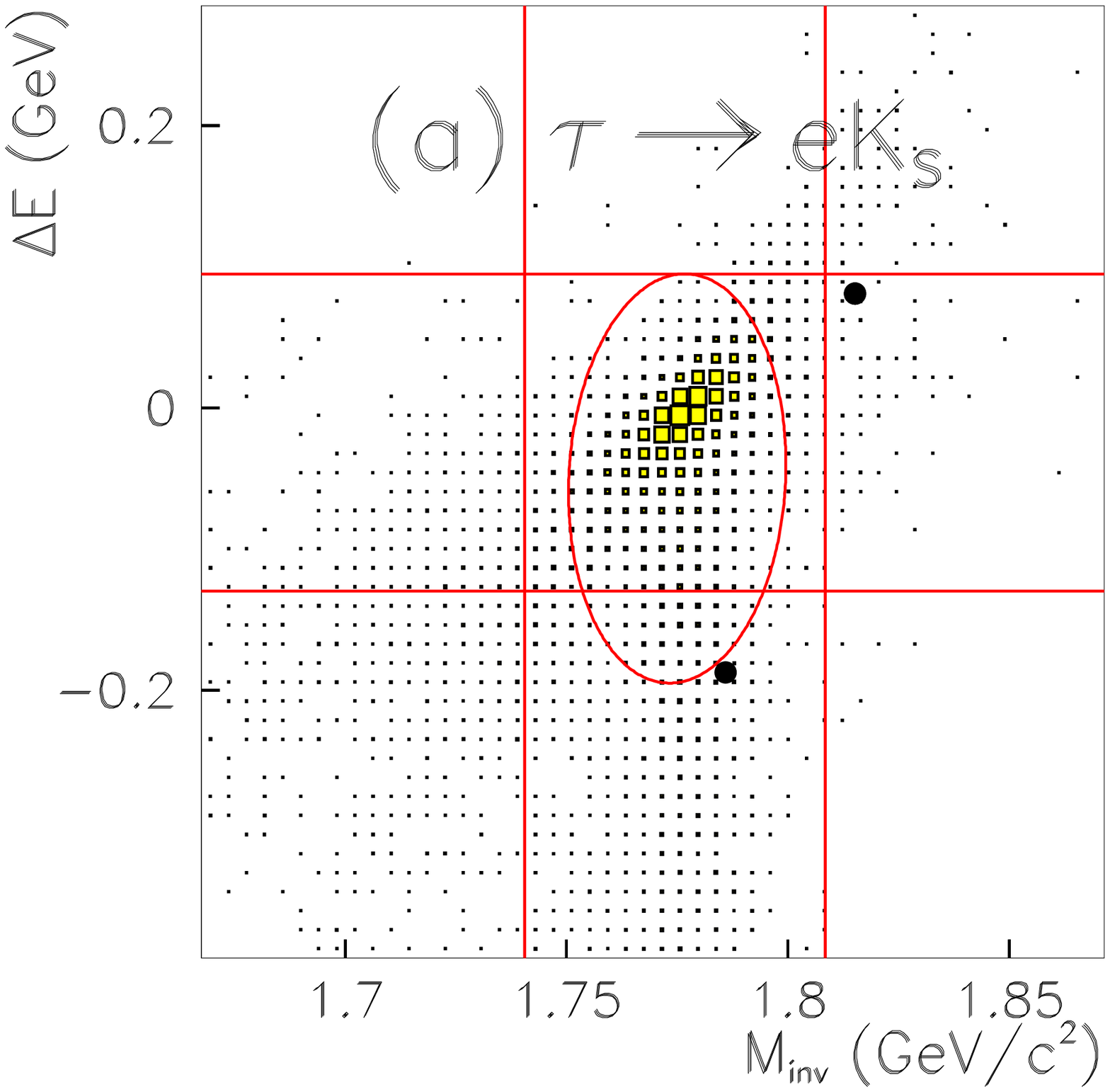}}
 \caption{
{Scatter-plots} of data in the $M_{\rm inv}$ -- $\Delta{E}$
plane: (a) and (b) correspond to
the $\pm 15 \sigma$ area for
the $\tau^-\rightarrow e^-\ks$ and $\tau^-\rightarrow \mu^-\ks$ modes, respectively.
{The elliptical signal region} shown by a solid curve in (a) and (b) 
is used for evaluating the signal yield.
In (a) and (b), the vertical  and horizontal lines denote 
$\pm5 \sigma$.
Closed circles correspond to the data.
The filled boxes show the MC signal distribution
with arbitrary normalization.
\label{fig:5}
}
\end{center}
\end{figure}

We blind a region of $\pm 5\sigma_{\rm{M_{inv}}}$ 
around the $\tau$ mass in $\rm{M_{\rm inv}}$ 
and 
a region of
$-0.5\mbox{ GeV} < \Delta E < 0.5$ GeV
{so as not to bias our choice of selection criteria.}
Figure~\ref{fig:5} shows scatter-plots 
for data and signal MC samples 
distributed over $\pm 15\sigma$ 
in the $M_{\rm{inv}}-\Delta E$ plane.
{Most of the surviving}
background events in both modes
come from 
$D^{\pm}\rightarrow\pi^{\pm}\ks$
and 
$D^{\pm}\rightarrow\ell^{\pm}\nu\ks$ decays.
The remaining continuum backgrounds in the $\tau^-\rightarrow \mu^-\ks$  mode
are combinations of a true $\ks$ meson and a fake lepton.

{To optimize the sensitivity for the search, we use an
elliptically shaped signal region, which has major and minor axes
that correspond to
{$\pm5 \sigma$ 
in the MC resolution for 
the $M_{\rm inv}-\Delta E$ plane.} 
This region is shown in  Fig.~\ref{fig:5} (a) and (b). 
{Signal efficiencies for this region} 
after all requirements are 11.8\% for 
$\tau^-\rightarrow e^-\ks$ 
and 13.5\% for 
$\tau^-\rightarrow \mu^-\ks$,
respectively.}


{As there are few remaining MC background events 
{in the signal ellipse,} 
we estimate the background contribution using the $M_{\rm{inv}}$  sideband regions.
This is achieved by extrapolating to the signal region under the
assumption that the background distribution is flat in $M_{\rm{inv}}$.
{We find}
the expected background in the ellipse 
to be $0.2 \pm 0.2$ events for each of the two modes.}
We open the blinded region and find
no data events for {the}
$\tau^-\rightarrow e^-\ks$ and $\tau^-\rightarrow \mu^-\ks$ modes
(see Fig.~\ref{fig:5} (a) and (b)). 
Since no statistically significant excess of data over
the expected background in the signal region {is} observed,
we apply a frequentist approach 
to calculate upper limits for the signal {yields}~\cite{cite:FC}.
The resulting limits for 
the signal yields at 90\% confidence level, $s_{90}$,  
are 
{2.23 events} 
in both modes.
The upper limits on the branching fraction
before the inclusion of systematic uncertainties are then
calculated as
\begin{equation}
{\cal B}(\tau \rightarrow \ell^- \ks) 
<  \frac{s_{90}}{2 \varepsilon N_{\tau\tau}{\cal B}
( \ks \rightarrow \pi^+\pi^-)}
\end{equation}
where $N_{\tau\tau} = 250 \times 10^6$ and 
${\cal B}(\ks \rightarrow \pi^+\pi^-) = 0.6895$~\cite{PDG}. 
The resulting values are
${\cal B}(\tau^-\rightarrow e^-\ks) < 5.5\times 10^{-8}$
and 
${\cal B}(\tau^-\rightarrow \mu^-\ks) < 4.8 \times 10^{-8}$. 

%
%

The dominant systematic uncertainties 
on 
{the detection sensitivity:} 
$2\varepsilon N_{\tau\tau}{\cal B}(\ks\rightarrow \pi^+\pi^-)$ 
{come
from $\ks$ reconstruction
and
tracking efficiencies.}
{These are 4.5\% and 4.0\%,
respectively, for both modes.}
Other sources of the systematic uncertainties
are:
the trigger efficiency (0.5\%), 
lepton identification (2.0\%),
{MC statistics (0.3\%),} and luminosity (1.4\%). 
Assuming no correlation between them,
all these uncertainties are combined in quadrature to 
give a total of {$6.5\%$}.  

{Upper limits on the branching fractions at the 90\% C.L.
including these systematic uncertainties are calculated} 
{with} the POLE program without conditioning
~\cite{cite:pole}.
The resulting upper limits on the branching fractions are
\begin{eqnarray*}
&&{\cal B}(\tau^-\rightarrow e^-\ks) < 5.6 \times 10^{-8} \\
&&{\cal B}(\tau^-\rightarrow \mu^-\ks) < 4.9 \times 10^{-8}.
\end{eqnarray*}

\section{Summary}

{In conclusion,}
we have searched for 
{the lepton flavor violation decays}
$\tau^-\rightarrow\ell^-\ks$ ($\ell = e \mbox{ or } \mu$)  
using data collected 
{with} the Belle detector at the KEKB $e^+e^-$ asymmetric-energy collider.
We found no signal in either mode.
The following  upper limits on
the branching fractions were obtained:
${\cal{B}}(\tau^-\rightarrow e^-\ks) < 5.6\times 10^{-8}$ 
and 
${\cal{B}}(\tau^-\rightarrow \mu^- \ks) < 4.9\times 10^{-8}$ 
at the 90\% confidence level
{and including systematic uncertainties}. 
{These results 
improve 
the search sensitivity
by factors of 16 and 19, respectively,
compared 
{to the} previous limits obtained by CLEO experiment.}

\section*{Acknowledgments}
We thank the KEKB group for the excellent operation of the
accelerator, the KEK cryogenics group for the efficient
operation of the solenoid, and the KEK computer group and
the National Institute of Informatics for valuable computing
and Super-SINET network support. We acknowledge support from
the Ministry of Education, Culture, Sports, Science, and
Technology of Japan and the Japan Society for the Promotion
of Science; the Australian Research Council and the
Australian Department of Education, Science and Training;
the National Science Foundation of China under contract
No.~10175071; the Department of Science and Technology of
India; the BK21 program of the Ministry of Education of
Korea and the CHEP SRC program of the Korea Science and
Engineering Foundation; the Polish State Committee for
Scientific Research under contract No.~2P03B 01324; the
Ministry of Science and Technology of the Russian
Federation; the Ministry of Education, Science and Sport of
the Republic of Slovenia; the National Science Council and
the Ministry of Education of Taiwan; and the U.S.\
Department of Energy.

\end{document}